%Paper: hep-th/9305172
%From: percacci@tsmi19.sissa.it
%Date: Sat, 29 May 1993 10:22:17 +0100

\def\thetacl{\theta_{\rm (cl)}{}}

\def\btheta{\bar\theta}
\def\bnabla{\bar\nabla}
\def\ba{\bar A}

\def\igamma{{\mit\Gamma}}
\font\titlefont=cmbx10 scaled\magstep1
\magnification=\magstep1
\null
\rightline{SISSA 71/93/EP}
\rightline{hep-th/9305172}
\vskip 1.5cm
\centerline{\titlefont AVERAGE EFFECTIVE POTENTIAL}
\centerline{\titlefont FOR THE CONFORMAL FACTOR}
\smallskip
\vskip 1.5cm
\centerline{\bf R. Floreanini \footnote{$^*$}{\tt florean@ts.infn.it}}
\smallskip
\centerline{Istituto Nazionale di Fisica Nucleare, Sezione di Trieste}
\centerline{Dipartimento di Fisica Teorica, Universit\`a di Trieste}
\centerline{Strada Costiera 11, 34014 Trieste, Italy}
\bigskip\smallskip
\centerline{\bf R. Percacci \footnote{$^{**}$}{\tt percacci@tsmi19.sissa.it}}
\smallskip
\centerline{International School for Advanced Studies, Trieste, Italy}
\centerline{via Beirut 4, 34014 Trieste, Italy}
\centerline{and}
\centerline{Istituto Nazionale di Fisica Nucleare, Sezione di Trieste}
\vskip 1.8cm
\centerline{\bf Abstract}
\smallskip\midinsert\narrower\narrower\noindent
In a four dimensional theory of gravity with lagrangian quadratic in
curvature and torsion, we compute the effective action for metrics
of the form $g_{\mu\nu}=\rho^2\delta_{\mu\nu}$, with $\rho$ constant.
Using standard field-theoretic methods we find that one loop quantum
effects produce a nontrivial effective potential for $\rho$.
We explain this unexpected result by showing how our regularization
procedure differs from the one that is usually adopted in Quantum Gravity.
Using the method of the average effective potential, we
compute the scale dependence of the v.e.v. of the conformal factor.
\endinsert
\vskip 1cm
\vfil\eject
In quantum field theory the vacuum expectation value (v.e.v.)
of the fields is usually determined by the effective potential
(the nonderivative part of the effective action).
In classical theories of gravity the possible form of a potential for
the metric is severely constrained by general covariance:
the only allowed local term in the Lagrangian depending on the metric
but not on its derivatives is the cosmological term.
We have suggested elsewhere that in Quantum Gravity the v.e.v.
of the metric could be fixed by an effective potential [1].
The particular dynamics that we employed there was based on
a bimetric Lagrangian, which one could think of as a mean field
approximation to an ordinary gravitational Lagrangian quadratic
in curvature and torsion. We observed that in the presence of two
metrics one could obtain a genuine potential term whose minimum
fixes the v.e.v. of the metric.

One could think that this result was due to the unconventional
dynamics that we started with. The main point we want to make in
this note is that the same result can be obtained starting from
an ordinary Lagrangian quadratic in curvature and torsion and
using the familiar background field method.
We will restrict our attention to the conformal sector and write
$$
g_{\mu\nu}=\rho^2\gamma_{\mu\nu}\eqno(1)
$$
where $\gamma_{\mu\nu}$ is a fixed fiducial metric.
In order to simplify the discussion as much as possible we will present
calculations only for the case $\gamma_{\mu\nu}=\delta_{\mu\nu}$,
but our results hold more generally.
The effective dynamics of the conformal factor $\rho$ induced by the
conformal anomaly of matter fields has been the subject of recent
investigations [2,3]. In this work we will discuss the effective
potential for $\rho$ in the framework of a gauge theory of gravity.

{}From standard Quantum Gravity arguments, one would expect
to find only a cosmological term, {\it i.e.} a potential proportional
to $\rho^4$. Instead, we find an effective potential of the
Coleman--Weinberg form, with the minimum occurring for nonzero $\rho$.
We will explain the origin of this result: it lies in the way in
which the regularization is defined.

We then discuss the renormalization group flow of the minimum
of the potential. We do this by computing the average effective
potential for $\rho$. The average effective action is a continuum
version of the block-spin action of lattice theories, which has been recently
applied to scalar and gauge theories [4,5].
We find that the v.e.v. of $\rho^2$ (and therefore of the metric)
is essentially constant up to Planck's energy, and scales according to
its canonical dimension (mass squared) above Planck's energy,
up to logarithmic corrections.
In the conclusion we offer some speculations on the physical meaning
of this behavior.

In the model we shall consider, the independent dynamical variables
are the vierbein $\theta^a{}_\mu$ and an $O(4)$ gauge field
$A_\mu{}^a{}_b$ (we shall concentrate on the Euclidean theory,
where $a,b=1,2,3,4$ are internal
indices and $\mu,\,\nu=1,2,3,4$ are spacetime indices).
With $\theta$ and $A$ we can construct metric, curvature and
torsion fields:
$$
\eqalignno{
&g_{\mu\nu}=\,\theta^a{}_\mu\, \theta^b{}_\nu\, \delta_{ab}\ ,&(2a)\cr
&F_{\mu\nu}{}^a{}_b=\,\partial_\mu A_\nu{}^a{}_b-\partial_\nu A_\mu{}^a{}_b
+eA_\mu{}^a{}_c A_\nu{}^c{}_b-eA_\nu{}^a{}_c A_\mu{}^c{}_b\ ,&(2b)\cr
&\Theta_\mu{}^a{}_\nu=\,\partial_\mu\theta^a{}_\nu-
\partial_\nu\theta^a{}_\mu
+eA_\mu{}^a{}_b\theta^b{}_\nu-eA_\nu{}^a{}_b\theta^b{}_\mu\ ,&(2c)\cr}
$$
where $e$ is the gauge coupling constant. As an action we take
$$
S(\theta,A)={1\over4}\int d^4x\ \sqrt{|\det g|}\,g^{\mu\rho}g^{\nu\sigma}
\left[\delta_{ac}\delta^{bd}
F_{\mu\nu}{}^a{}_b \, F_{\rho\sigma}{}^c{}_d \
+\delta_{ab}\Theta_\mu{}^a{}_\nu\Theta_\rho{}^b{}_\sigma \right].
\eqno(3)
$$
It is manifestly invariant under local $O(4)$ and general coordinate
transformations. Note that $\theta^a{}_\mu$ and $A_{\mu ab}$ have
canonical dimension of mass, and $g_{\mu\nu}$ of mass squared.
In (1) we take $\rho$ to carry dimension of mass.
Let us note right away that $g_{\mu\nu}$ can not be the ``geometric''
metric, which has to be dimensionless (since we are
assuming that the coordinates have dimensions of length).
We shall return to this point later.

We will evaluate the one-loop effective potential for the conformal
factor $\rho$ using the background field method.
We first expand $S$ up to second order around the classical solution
of the field equations $A_{\rm (cl)\mu}{}^a{}_b=0$,
$\thetacl^a{}_\mu=\rho\, \delta^a_\mu$, with $\rho$ constant.
The linearized action has the form
$$\eqalign{S^{(2)}=\
&{1\over2}\int d^4x \Big[\delta A_{\mu ab} \delta^{ac}
\bigl(\delta^{bd}
\left(-\delta^{\mu\rho}\partial^2+\partial^\mu\partial^\rho\right)
+e^2\rho^2\left(\delta^{bd}\delta^{\mu\rho}-\delta^{b\rho}\delta^{d\mu}\right)
\bigr)\delta A_{\rho cd}\cr
&-2e\rho\,\delta\theta^a{}_\mu\left(\delta^{d\mu}\partial^\rho-
\delta^{\mu\rho}\partial^d\right)\delta^{ac}\delta A_{\rho cd}
+\delta\theta^a{}_\mu\delta_{ac}\left(-\delta^{\mu\rho}\partial^2
+\partial^\mu\partial^\rho\right)\delta\theta^c{}_\rho \Big]\ .\cr}
\eqno(4)
$$
In this expression, indices are raised and lowered with $\delta_{\mu\nu}$.
This linearized action is invariant under the linearized gauge
transformations and linearized coordinate transformations.
We add to the linearized action the gauge-fixing terms
$$
\int d^4x \left[{1\over2\alpha}(\partial_\mu\delta A^\mu_{ab})^2
+{1\over2\beta}(\partial_\mu\delta\theta^{a\mu})^2\right]\ .
\eqno(5)
$$
The effective action is one half the logarithm of the determinant
of the differential operator appearing in (4), taking into account
gauge fixing and ghost terms. The operator can be
diagonalized using the method of the spin projectors, which is discussed
for example in [6]. More details will be given elsewhere. We find
$$
\Gamma(\rho)={1\over2}\int d^4x \int {d^4q\over(2\pi)^4}
\left[(5+3)\ln(q^2+{1\over2}e^2\rho^2)+
3\ln(q^2+e^2\rho^2)+\ln(q^2+2e^2\rho^2)\right]\ \eqno(6)
$$
plus terms independent of $\rho$ (we used the notation
$q^2=\delta^{\mu\nu}q_\mu q_\nu$). The first term comes from the
modes with spin $2^-$ and $1^-$, the second from those with spin $1^+$,
the last from those with spin $0^-$. The ghost contribution turns out to be
independent of $\rho$.

The integral can be regularized with a simple cutoff $\Lambda$.
Adding suitable counterterms of the form $\Lambda^2\rho^2$
and $\rho^4\ln\Lambda$ one arrives at the renormalized effective
potential
$$
\eqalignno{
&\Gamma_{\rm ren}(\rho,\gamma)=
\int d^4x\,\sqrt{\gamma}\,V_0(\rho)\ ,&(7a)\cr
&V_0(\rho)={9\over64\pi^2}
e^4\rho^4\left(\ln\left({e^2\rho^2\over
\mu^2}\right)-{1\over2}\right)\ ,&(7b)\cr}
$$
where $\mu$ is a renormalization constant with dimensions of mass;
we have written the result for an arbitrary constant $\gamma_{\mu\nu}$.
This potential has the same form of the one we computed previously
in the mean field approach [1]. It has a minimum for
$\rho=\rho_0=\mu/e$.

The potential (7) is not simply a cosmological constant.
This result is surprising. The classical theory depends on $\rho$
and $\gamma_{\mu\nu}$ only through the combination $g_{\mu\nu}$ given in
(1). This gives rise to invariance under the Weyl transformations
$$
\gamma'_{\mu\nu}=\omega^2\gamma_{\mu\nu}\ ,\qquad
\qquad\rho'=\omega^{-1}\rho\ .\eqno(8)
$$
The theory can be quantized in such a way that this symmetry is
preserved [7]. As a consequence, also the quantum effective action
should depend on $\gamma_{\mu\nu}$ and $\rho$ only through the
combination $g_{\mu\nu}$, and this is not the case for (7).
It is the regularization procedure that we have chosen that
breaks this invariance. In fact we have integrated over the range
of momenta $\gamma^{\mu\nu}q_\mu q_\nu<\Lambda^2$;  this introduces
a dependence of the theory on $\gamma_{\mu\nu}$ alone, not accompanied by a
factor of $\rho$, and is ultimately responsible for the appearance
of the logarithm of $\rho$ in (7).
Note that one could add to $\Gamma(\rho,\gamma)$ the local counterterm
$(9/64\pi^2)\int d^4x\,\sqrt{\gamma}\,e^4\rho^4\ln(\gamma^{1/4})$.
This would restore the invariance under the transformations (8)
but would break diffeomorphisms. In fact,
our effective action is invariant under diffeomorphisms if
$g_{\mu\nu}$ and $\gamma_{\mu\nu}$ are transformed simultaneously,
and $\rho$ is treated as a scalar field.

There is an alternative way of regulating the theory: integrate over
the range of momenta
$g^{\mu\nu}q_\mu q_\nu=\rho^{-2}\gamma^{\mu\nu}q_\mu q_\nu<\lambda^2$,
with $\lambda$ a dimensionless cutoff. Redefining the integration
variables as $q'_\mu=\rho^{-1} q_\mu$ and discarding a term proportional
to $\delta(0)$, the integrals in (6) are reduced to the general form
$\rho^4\int d^4 q'\ \ln(\gamma^{\mu\nu}q'_\mu q'_\nu+c)$,
with $c$ a dimensionless constant independent of $\rho$,
the integration being now over the range
$\gamma^{\mu\nu}q'_\mu q'_\nu<\lambda^2$.
The important point is that the integral does not depent on $\rho$
anymore. Thus after renormalization, the effective action would be of the
form $\Gamma\sim\int d^4x\,\sqrt{\gamma}\,\rho^4=\int d^4x\,\sqrt{g}$,
{\it i.e.} a cosmological term.
{}From the point of view of quantum field theory it is
unusual to have a regularization which itself
depends on the dynamical variable. Nevertheless, this is the choice
which is tacitly made in most works on Quantum Gravity.

The difference between the two ways of implementing the cutoff is that
in the former case the domain of integration is independent on the
dynamical variable $\theta^a{}_\mu$, while in the latter it depends on it.
Both procedures are mathematically correct and the choice between the
two has to be dictated by physical arguments.
One could argue that the correct quantization is the one that preserves
the Weyl invariance (8), but in the present case it seems that breaking
this invariance would not violate any physical principle.
Since forming the modulus squared of the four-momentum is a geometric
construction, the distinction between the two procedures has to do
with what metric is taken to represent the geometry of spacetime.
As already remarked, the geometry cannot be given directly by the
composite operator $g_{\mu\nu}$, since it is dimensionful.
It will be related to it by a constant of proportionality $\ell^2$
having dimension of length squared.

If the geometry is given by the metric $\ell^2\, g_{\mu\nu}$,
the cutoff depends on the metric and the effective potential
will be just a cosmological term, as shown above.
In this case the nondegeneracy of the metric has to be imposed from
the outside, and the parameter $\ell$ is undetermined.
On the other hand the geometry could be given by the metric
$\ell^2\, \langle g_{\mu\nu}\rangle
=\ell^2\langle\rho^2\rangle\gamma_{\mu\nu}=
\ell^2\,\rho_0^2\,\gamma_{\mu\nu}$.
In this case it is natural to identify $\ell^{-1}=\rho_0$,
in which case the geometric metric coincides with $\gamma_{\mu\nu}$.
This is the point of view implicit in our calculations,
and it leads to the effective potential (7). In this approach the
nondegeneracy of $g_{\mu\nu}$ is a result of the quantum dynamics of
the theory. It also has the advantage that it does not
necessitate the introduction of an external dimensionful parameter.
We note here that the presence of a nontrivial effective potential for the
conformal factor is also relevant to the problem of the cosmological
constant [8].

We turn now to another definition of the effective action
which has been applied recently to scalar and gauge field theories:
the so-called average effective action [4,5].
This will allow us to compute the scale dependence of the
effective potential, and hence of the v.e.v. of the operator
$g_{\mu\nu}$.
The average effective action depends on a momentum scale $k$.
To define it, one begins by adding to the action (3)
quadratic terms which constrain the averages of the fields $\theta$
and $A$ in volumes of size $k^{-4}$ centered around the point $x$
to take certain values $\btheta(x)$ and $\ba(x)$ (up to small
fluctuations):
$$
\eqalign{S_{\rm constr}=
\int d^4x\ \sqrt{\gamma}
\Big[&{1\over4}(\bar F_{\mu\nu ab}-f_k F_{\mu\nu ab})\,
{\gamma^{\mu\rho}\gamma^{\nu\sigma}\over 1-f_k^2}\,
(\bar F_{\rho\sigma}{}^{ab}-f_k F_{\rho\sigma}{}^{ab})\cr
+&{1\over2\alpha}\gamma^{\mu\nu}\bnabla_\mu(A-\ba)_{\nu ab}\,
{1\over 1-f_k^2}\,
\gamma^{\rho\sigma}\bnabla_\rho(A-\ba)_{\sigma}{}^{ab}\cr
+&{1\over2}\bnabla_\mu(\btheta-f_k\theta)^a{}_\nu\,
{\gamma^{\mu\rho}\gamma^{\nu\sigma}\over 1-f_k^2}\,
\bnabla_\rho(\btheta-f_k\theta)^a{}_\sigma\Big]\ .
\cr} \eqno(9)
$$
In this formula $f_k=f_k(-\gamma^{\mu\nu}\bnabla_\mu\bnabla_\nu)$,
where
$\bnabla_\mu\theta^a{}_\nu=\partial_\mu\theta^a{}_\nu
+e\ba_\mu{}^a{}_b\theta^b{}_\nu
-\igamma^\lambda_{\mu\nu}\theta^a{}_\lambda$
and $\igamma$ are the Christoffel symbols for the metric
$\gamma_{\mu\nu}$. The differential operator
$f_k(-\gamma^{\mu\nu}\bnabla_\mu\bnabla_\nu)$
will perform the desired averaging operation
if we take $f_k(x)=exp(-a(x/k^2)^b)$, with $a$, $b$ constant parameters.
Note that the explicit introduction of the fields $\btheta$ and $\ba$
breaks both coordinate and gauge invariance, so no further gauge
fixing is needed.
In (9) we have contracted all spacetime indices with
the metric $\gamma_{\mu\nu}$, in line with our assumption that
it is this metric that dictates the geometry. Other choices are
possible but will not be considered here.

In order to compute the average effective potential we choose the
average fields $\ba=0$ and $\btheta^a{}_\mu=\rho\,\delta^a_\mu$
with $\rho$ constant.
If the parameter $b$ in $f_k$ is chosen larger than 2, the Ansatz
$A_{{\rm cl}\mu ab}=\ba_{\mu ab}$ and
$\theta_{\rm cl}{}^a{}_\mu=\btheta^a{}_\mu$
gives a solution of the classical equations of motion of the total action.
Proceeding as before, we arrive at the average action
$$
\eqalign{
\Gamma_k(\rho)={1\over2}\int d^4x \int {d^4q\over(2\pi)^4}
\Bigl[&8\ln\left(P_k+{1\over2}e^2\rho^2\right)
+6\ln\left(P_k+{1\over2}e^2\rho^2 f_k^2\right)\cr
&+3\ln\left(P_k^2+e^2\rho^2P_k\left(1+\left(\alpha+{1\over2}\right)f_k^2\right)
+\alpha e^4\rho^4f_k^2\right)\cr
& +3\ln\left(P_k+{1\over2}\alpha e^2\rho^2f_k^2\right)
+\ln\left(P_k+2e^2\rho^2\right)\Bigr]\cr}
\eqno(10)
$$
where $f_k^2=\big(f_k(q^2)\big)^2$ and $P_k(q^2)=q^2/(1-f_k^2)$.
Note that in the limit $k\rightarrow 0$,
the function $f_k$ becomes zero and $P_k$ becomes equal
to $q^2$. One can then easily check that up to field-independent terms,
$\Gamma_0=\Gamma_{k=0}$ reduces to the old effective action (6).
One can split
$\Gamma_k(\rho)=\Gamma_0(\rho)+\Delta\Gamma_k(\rho)$,
where $\Gamma_0(\rho)$ contains the divergences but is independent
of $k$ and
$$
\eqalign{
\Delta\Gamma_k(\rho)={1\over2}\int d^4x \int {d^4q\over(2\pi)^4}
\Biggl[&
8\ln\left({P_k+{1\over2}e^2\rho^2\over
q^2+{1\over2}e^2\rho^2}\right)
+6\ln\left({P_k+{1\over2}e^2\rho^2 f_k^2\over q^2}\right)\cr
& +3\ln\left(
{P_k^2+e^2\rho^2P_k\left(1+\left(\alpha+{1\over2}\right)f_k^2\right)
+\alpha e^4\rho^4f_k^2  \over q^2(q^2+e^2\rho^2)}\right)\cr
& +3\ln\left({P_k+{1\over2}\alpha e^2\rho^2f_k^2\over q^2}\right)
+\ln\left({P_k+2e^2\rho^2\over q^2+2 e^2\rho^2}\right)\Biggr]}
\eqno(11)
$$
is automatically convergent. The part $\Gamma_0(\rho)$ can be
renormalized as before, leading to the effective potential
$V_0(\rho)$ given in (7). Define the average effective potential
$V_k(\rho)=V_0(\rho)+\Delta V_k(\rho)$ by
$\Gamma_k(\rho)=\int d^4x\,\sqrt{\gamma}\,V_k(\rho)$.
To find its minimum we have to solve the equation
$$
0={\partial V_k(\rho)\over\partial(\rho^2)}=
{e^2\over 64\pi^2}\left[18e^2\rho^2\ln{e^2\rho^2\over\mu^2}
+k^2F\left({e^2\rho^2\over k^2}\right)\right]\ ,\eqno(12)
$$
where, using the dimensionless variables $x=q^2/k^2$,
$t=e^2\rho^2/k^2$ and $\tilde P(x)=P_k(q^2)/k^2=x/(1+f^2)$,
$f^2=\exp(-2ax^b)$, the function $F$ is given by
$$
\eqalign{
F(t)={64\pi^2\over e^2k^2}
{\partial \Delta V_k(\rho)\over\partial(\rho^2)}=
&\,2\int_0^\infty\! dx\,x\, f^2
\Biggl[-{4\tilde P\over (\tilde P+{t\over2})(x+{t\over2})}
+{3\over \tilde P+{1\over2}tf^2}\cr
& +3{\tilde P({1\over2}x-\tilde P)+\alpha(\tilde Px+t(2x+t))
\over (x+t)\bigl(\tilde P^2+\tilde P t
(1+(\alpha+{1\over2})f^2)+\alpha t^2f^2\bigr)}\cr
&+{3\over2}{\alpha\over \tilde P+{\alpha\over2}tf^2}
-{2\tilde P\over(\tilde P+2t)(x+2t)}
\Biggr]\ .\cr}
\eqno(13)
$$
This function can be studied numerically. Choosing $a=1$, $b=3.19$ in
$f^2$, (see [4]) and setting $\alpha=0$, $F(t)$ grows from
$F(0)=-c_1\sim-12$ to zero for $t\sim5$, it reaches a
maximum of order 0.2 for $t\sim 15$ and decreases slowly to zero for
large $t$ like $K/t$ for $K$ slowly varying.
The minimum of the effective potential can be plotted
numerically. One can only study analytically the behavior for
$t$ very large and very small. Let us denote $\rho_k$
the minimum of $V_k$. For $k=0$,
$\rho_0=\mu/e$. For $t\gg1$ (which corresponds to $k\ll\mu$)
we can expand $\rho_k=\rho_0+\epsilon$,
and use the asymptotic behavior $F(t)\sim K/t$.
Inserting in (12) one finds
$$
\rho^2_k=\rho_0^2
\left(1-{K\over18}{k^4\over\mu^4}\right)\ .\eqno(14)
$$
On the other hand for $t\ll 1$ we can expand the function $F$ in
Taylor series around $t=0$: $F(t)=-c_1+c_2 t+\ldots$.
Equation (12) shows that $\rho^2_k$ grows
slower than $c_1 k^2$ and faster than
$c_1k^2/\left(18\ln({c_1 k^2/\mu^2})+c_2\right)$.
For $k\gg\mu$ the denominator becomes large and this justifies
a posteriori the approximation $t\ll1$. In fact, this can also be
checked numerically.

For $\alpha\not =0$ but not too large, the behavior of the
potential is essentially the same.
We note that the behavior for large and small $k$ agrees with the one
found using a mean field approximation and treating $k$ simply as a
sharp infrared cutoff [9].

We now return to the discussion of the physical implications
of our results. The discussion on the choice of the geometric metric
given earlier for the case $k=0$ can be repeated for $k>0$.
Assuming that the geometry is given by $\ell^2\langle g_{\mu\nu}\rangle$
one has the further option of identifying $\ell^{-1}$ with $\rho_0$ or
$\rho_k$. In the first case the geometry is given by the $k$-dependent metric
$(\rho_k^2/\rho_0^2)\gamma_{\mu\nu}$,
while in the second the geometry is fixed, given by $\gamma_{\mu\nu}$.
(This is the choice that was made in writing (9).)
Either way, the scale dependence of the metric can lead to
striking effects.
For example, in the case of a scalar field coupled minimally to the
$k$-dependent geometric metric, the scaling of the metric
improves the ultraviolet behavior of the propagator and shifts the
physical pole [9].

Here we shall briefly discuss an alternative approach, in which a
(dimensionless) scalar field is coupled directly to the quantum
field $g_{\mu\nu}$ [10]:
$$
S_{\rm scalar}(\varphi)={1\over2}\int d^4x\ \sqrt{g}
\left(g^{\mu\nu}\partial_\mu\varphi\partial_\nu\varphi
+c^2\varphi^2\right)\ .
\eqno(15)
$$
Defining the canonical field $\phi=\rho\varphi$
the action can be rewritten in the form
$$
S_{\rm scalar}(\phi)={1\over2}\int d^4x\ \sqrt{\gamma}
\left(\gamma^{\mu\nu}\partial_\mu\phi\partial_\nu\phi
+c^2\rho^2\phi^2+\ldots\right)\ ,
\eqno(16)
$$
where the dots represent terms containing derivatives of $\rho$.
When $\rho$ is constant, the Lagrangian of $\phi$ has the same general
form of the linearized Lagrangian (4) and therefore quantum fluctuations
of $\phi$ contribute to the effective potential for $\rho$
(see also [3]).

On the other hand, the running of $\rho$ directly affects the
propagator of $\phi$. We assume that in the propagator for a free
particle of four-momentum $q_\mu$ the ``mass''
$c\langle\rho\rangle$ has to be taken at scale $k=|q|$
(a similar assumption was discussed recently in a different context [11]).
The inverse propagator of $\phi$ would then have the form
$q^2+c^2\rho^2_q$, with $\rho^2_q$ approximately constant
for $q^2<\rho_0^2$ and growing roughly like $q^2$ for
$q^2>\rho_0^2$. The physical pole of the propagator occurs at mass
approximately equal to $c\rho_0$ for $c<1$, but is shifted to
exponentially large values for $c>1$. In fact, a positive
anomalous dimension for $\rho$ could make the pole disappear
altogether. The mass $\rho_0$ has to be identified with
Planck's mass [1,10]. Thus, particles with
masses larger than Planck's mass would essentially disappear from the
spectrum. One may hope that a mechanism of this type is capable of
removing the ghosts of the gravitational sector. This seems to be
a restatement of the criterion given in [12].

To summarize, we have suggested that there exists an alternative
method for quantizing a gauge theory of gravity which produces a
nontrivial effective potential for the metric.
The difference from the traditional approach lies therein, that the
domain of integration over the momenta is determined by some fixed
metric, and not by the dynamical metric. In this sense it is a
bimetric theory, even though at the level of the starting classical
Lagrangian no second metric appears.
In this it differs from the mean-field approach of [1].
General covariance will be preserved provided both metrics are
transformed simultaneously, although this aspect cannot be completely
appreciated by looking just at the effective potential.

Concerning the scale dependence of the metric, one should perform
a more sophisticated analysis by taking into account the running
of the coupling constants, along the lines of [5].
Nevertheless we believe that our simple minded approach is sufficient
to capture at least some qualitative features of this phenomenon.
We plan to return on these open problems in the future.

\vskip 1 truecm
\centerline{\bf Acknowledgements}
\noindent We benefitted from discussions with
D. Anselmi, S. Bellucci, M. Fabbrichesi, L. Griguolo, M. Reuter,
A. Schwimmer, E. Spallucci, K.S. Stelle and M. Tonin.
\vskip 1 truecm

\centerline{\bf References}
\bigskip
\noindent
\item{1.} R. Floreanini, E. Spallucci and R. Percacci, Class. and
Quantum Grav. {\bf 8}, L193 (1991);
R. Floreanini and R. Percacci, Phys. Rev. D {\bf 46}, 1566 (1992).
\smallskip
\item{2.} I. Antoniadis and E. Mottola, Phys. Rev. D {\bf 45}, 2013
(1992);\hfil\break
I. Antoniadis, P. Mazur and E. Mottola, Nucl. Phys. {\bf B 388}, 627
(1992);\hfil\break
S.D. Odintsov, Z. Phys. C {\bf 54}, 531 (1992).
\smallskip
\item{3.} E. Elizalde and S.D. Odintsov, ``Gravitational phase
transitions in infrared quantum gravity'', Hiroshima preprint HUPD-92-10;
\hfil\break
A.A. Bytsenko, E. Elizalde and S.D. Odintsov, ``The renormalization
group and effective potential in curved spacetime with torsion'',
HUPD-93
\smallskip
\item{4.} C. Wetterich, Nucl. Phys. {\bf B 334}, 506 (1990);
{\it ibid.}{\bf B 352}, 529 (1991);
`` The average action for scalar fields near phase transitions'',
(to appear in Z. Phys. C).
\smallskip
\item{5.} M. Reuter and C. Wetterich, ``Average action for the Higgs
model with abelian gauge symmetry'', DESY 92-037 (to appear in Nucl.
Phys.); ``Running gauge coupling in three dimensions and the
electroweak phase transition'', DESY 93-006.
\smallskip
\item{6.} E. Sezgin and P. van Nieuwenhuizen, Phys. Rev. D {\bf 21},
3269 (1980).
\smallskip
\item{7.} N.C. Tsamis and R.P. Woodard, Ann. of Phys. {\bf 168}, 457
(1986).
\smallskip
\item{8.} E.T. Tomboulis, Nucl. Phys. {\bf B 329}, 410 (1990).
\smallskip
\item{9.} R. Percacci and J. Russo, Mod. Phys. Lett. A {\bf 7}, 865
(1992).
\smallskip
\item{10.} V. de Alfaro, S. Fubini and G. Furlan, Il Nuovo Cimento
{\bf A 50}, 523 (1979); {\it ibidem} {\bf B 57}, 227 (1980);
Phys. Lett. {\bf B 97}, 67 (1980).
\smallskip
\item{11.} S.B. Liao and J. Polonyi, ``Dynamical mass generation
without symmetry breaking'', MIT CTP 2136 (1992).
\smallskip
\item{12.} J. Julve and M. Tonin, Nuovo Cimento {\bf 46 B}, 137 (1978);
\hfil\break
A. Salam and J. Strathdee, Phys. Rev. D {\bf 18}, 4480 (1978).
\vfil
\eject
\bye